\documentclass[%
 reprint,
 superscriptaddress,
 amsmath,amssymb,
 aps,
]{revtex4-2}
\usepackage{hyperref}
\usepackage{graphicx}

\begin{document}

\title{Fast-update in self-learning algorithm for continuous-time quantum Monte Carlo}
\author{Ruixiao Cao}
\email{ruixiao.cao@phys.s.u-tokyo.ac.jp}
\affiliation{Department of Physics, University of Tokyo, Tokyo, 113-0033, Japan}

\author{Synge Todo}
\email{wistaria@phys.s.u-tokyo.ac.jp}
\affiliation{Department of Physics, University of Tokyo, Tokyo, 113-0033, Japan}
\affiliation{Institute for Physics of Intelligence, University of Tokyo, 113-0033, Japan}
\affiliation{Institute for Solid State Physics, University of Tokyo, Kashiwa, 277-8581, Japan}

\date{\today}

\begin{abstract}
We propose a novel technique for speeding up the self-learning Monte Carlo method applied to the single-site impurity model.
For the case where the effective Hamiltonian is expressed by polynomial functions of differences of imaginary-time coordinate between vertices, we can remove the dependence of CPU time on the number of vertices, $n$, by saving and updating some coefficients for each insertion and deletion process.
As a result, the total cost for a single-step update is drastically reduced from $O(nm)$ to  $O(m^2)$ with $m$ being the order of polynomials in the effective Hamiltonian.
Even for the existing algorithms, in which the absolute value is used instead of the difference as the variable of polynomial functions, we can limit the CPU time for a single step of Monte Carlo update to $O(m^2 + m \log n)$ with the help of balanced binary search trees.
We demonstrate that our proposed algorithm with only logarithmic $n$-dependence achieves an exponential speedup from the existing methods, which suffer from severe performance issues at low temperatures.
\end{abstract}

\maketitle

\section{Introduction}

Quantum Monte Carlo has been proved as one of the most powerful numerical methods on quantum many-body systems~\cite{LandauB2014, GubernatisKW2016}.
It has already become a large family of algorithms, but one of the common problems among these methods is the CPU time efficiency.
Although all Monte Carlo algorithms can give the exact measurement of any observables in the infinite Monte Carlo step limit, the statistical error is proportional to $1 / \sqrt{N}$ with $N$ being the Monte Carlo steps.
This property means that any speedup on Monte Carlo algorithms will automatically make a significant difference in its performance.

Here, we focus on a continuous-time quantum Monte Carlo algorithm~\cite{CT-AUX-1, CT-AUX0, CT-AUX1, CTAlgorithmsGull}, especially on its application to the single-impurity Anderson model~\cite{anderson1961localized}.
The model describes the interaction between fermions at a local impurity and the bath, formed by a set of modes of the pure system.
The Hamiltonian consists of non-interacting and interacting terms that can be written as~\cite{CTAlgorithmsGull}
\begin{align}
    H &= H_0 + H_1 \\
    H_0 &= -(\mu - U / 2)(n_{\uparrow} + n_\downarrow) + \sum_{\sigma, p}(V c_\sigma^\dagger a_{p, \sigma} + \text{h.c.}) \nonumber \\
    & \qquad + \sum_{\sigma, p} \epsilon_p a^\dagger_{p, \sigma}a_{p, \sigma} + K \\
    H_1 &= U(n_{\uparrow} n_{\downarrow} - (n_{\uparrow} + n_{\downarrow}) / 2) - K,
\end{align}
where $n_\sigma = c^\dagger_\sigma c_\sigma$ ($\sigma = \uparrow, \downarrow$) are the number operators of local fermions, $a_{p, \sigma}$ are the fermion operators for the bath, $\mu$ is the chemical potential, $U$ denotes the on-site Coulomb repulsion at the impurity site, and $V$ gives the hybridization between the impurity and the bath.
The parameter $K$ is an arbitrary positive constant of $O(1)$. One of the state-of-the-art algorithms on this model is the continuous-time auxiliary-field method (CT-AUX)~\cite{CT-AUX-1, CT-AUX0, CT-AUX1}, in which we introduce an auxiliary field $s$ for the interaction between upward and downward spin modes of the local impurity and rewrite $H_1$ as~\cite{nagai2017self, CTAlgorithmsGull, model0, model1}
\begin{equation}
    H_1 = -\frac{K}{2} \sum_{s = \pm 1} e^{\gamma s (n_\uparrow - n_\downarrow)},
\end{equation}
where $\gamma=\cosh^{-1} [1 + U/(2K) ]$.
Then, we can make a simulation by sampling over the spin vertices inserted into the imaginary-time axis. Namely, the configuration will be $\{(s_i, \tau_i)\}$, a set of spin vertices with their imaginary-time coordinates of auxiliary field. We can use various techniques to update the configuration, but the simplest is to either insert a new vertex at some imaginary time or remove an existing vertex in the configuration.

To decide whether to accept or reject a proposal of a new configuration, despite what update technique has been used, we need to calculate the weight difference between the initial and final states in addition to the possibility of making such a proposal.
Usually, the latter can be easily obtained. 
For example, in CT-AUX, if we consider the most straightforward update of insertion or removal of a single vertex $(s_i, \tau_i)$, the possibility of a proposal of removing an existing vertex is proportional to $1/n$, where $n$ is the number of vertices, and insertion on a given imaginary-time point will cause a factor proportional to the length of the imaginary-time axis, namely, the inverse temperature $\beta$.
As for the weight, for a single configuration, it can be calculated by evaluating a determinant of a square matrix of size $n$~\cite{CTAlgorithmsGull, CT-AUX1}.
As a result, the simplest brute-force Monte Carlo update will cost $O(n^3)$ CPU time to calculate the weight.
Fortunately, as the update of the configuration will only change the matrix of which we want to calculate the determinant locally, a fast-update technique can be applied to the Monte Carlo algorithm, and we can reduce the time complexity to $O(n^2)$ for each step~\cite{gullfastupdate}.
Still, calculating the weight is quite demanding computationally at low temperatures, as the average number of vertices increases in proportional to $\beta U$.

Recently, the self-learning Monte Carlo~(SLMC) technique~\cite{SLMC0, SLMC1, SLMC2} has been considered to be applied to CT-AUX~\cite{nagai2017self, nagai2020self}.
In this new method, an effective Hamiltonian is used for single update steps, the parameters of which are learned from the original Hamiltonian.
Instead of the $O(n^2)$ cost for calculating the weight difference of the original Hamiltonian, we should find an effective Hamiltonian where the weight can be calculated much more efficiently.
We run a number of Monte Carlo steps with the effective Hamiltonian to propose a configuration and finally return to the original Hamiltonian to decide whether to accept or reject it, to retain the detailed balance condition. 

One proposal~\cite{nagai2017self} of the effective Hamiltonian is to prepare such a Hamiltonian by training a linear regression model with simulation data. The effective Hamiltonian can be written as
\begin{align}
    -\beta H^{\text{eff}}_n(\{(s_i, \tau_i)\}) &= \frac{1}{n} \sum_{i, j} J(\tau_i - \tau_j) s_i s_j \nonumber \\
    & \qquad + \frac{1}{n} \sum_{i, j} L(\tau_i - \tau_j) + f(n).
    \label{eqn:SLMC0}
\end{align}
Here, $L$ and $J$ are linear combinations of Chebyshev polynomials of the absolute values of the imaginary time difference, coefficients of which can be trained with a set of $\{(s_i, \tau_i)\}$ with weight calculated exactly by the original Hamiltonian, namely using the fast-update of CT-AUX.
$f(n)$ is a function only depends on the number of vertices in the configuration and is also trained in a Chebyshev form.
Note that we assume that the imaginary time is normalized as $\tau_i \rightarrow \tau_i/\beta$ so that $\{(\tau_i - \tau_j)\}$ fits the interval $[-1,1]$.
Orders of each term are fixed in advance to some small numbers. For example, the orders are $m_{\textrm{c}, J} = m_{\textrm{c}, L} = 12$ for $L$ and $J$, and $m_{\textrm{c}, f} = 3$ for $f(n)$ in \cite{nagai2017self}.
Hereafter, we will use $m$ instead of $m_{\textrm{c}, L}$, $m_{\textrm{c}, J}$, and $m_{\textrm{c}, f}$ for simplicity, unless we need to distinguish between them.

Using the effective Hamiltonian, we can reduce the CPU time for a single Monte Carlo step down to $O(mn)$.
This cost is much smaller than the original $O(n^2)$ method.
After a number of Monte Carlo steps with the effective Hamiltonian, we calculate the weight with the original determinant algorithm (with CPU time of $O(n^3)$) and then accept or reject the proposed configuration. Thus, the total time cost for a single step is $O(n^3 + m n l)$, where $l$ is the length of updates with the effective Hamiltonian.
Ideally, we choose $l$ as $\tau_{\textrm{ori}}$, the auto-correlation time of the original update, so we need to compare the $O(n^2 \tau_{\textrm{ori}})$ complexity for the original Hamiltonian to generate one independent sample, with $O((n^3 + m n \tau_{\textrm{ori}})\tau_{\textrm{SL}})$, where $\tau_{\textrm{SL}}$ is the auto-correlation time for the new update, which should be small after a large number ($\tau_{\textrm{ori}}$) of proposals with the effective Hamiltonian. If $\tau_{\textrm{ori}}$ is long enough ($\tau_{\textrm{ori}} \gg n^2/m$), the speedup can be expressed as
\begin{equation}
    t_s \sim \frac{n}{m \tau_{\textrm{SL}}},
\end{equation}
and will be a good speedup for large enough $n$, or at low enough temperature, as well as $\tau_{\textrm{SL}}$ is not so long~\cite{nagai2017self}.

In the present paper, we propose a new technique for calculating the weight difference of the effective Hamiltonian in Eq.~(\ref{eqn:SLMC0}). We will show that the CPU time complexity for a single step can be reduced to $O(m^2 + m \log n)$, compared with $O(mn)$ of the original SLMC update.
Our method gives the same weight as the SLMC algorithm, but the $n$-dependence of the CPU time complexity has been reduced to logarithmic.
So we can expect an exponential speedup, especially at low temperature, where severe efficiency issues have been reported~\cite{yamamoto2018heat}. 

We should also notice that this form of effective Hamiltonian is free from the detailed structure of the models we are considering, and it fits for any cases where we can write the configuration by a set of $\{\tau_i\}$ of local degrees of freedom and imaginary times.
So the technique introduced below should be helpful for other models as well, e.g., boson models.

\section{Moment pre-calculation technique}
\label{sec:level1}

First, we consider the algorithm for the case where $J$ and $L$ are simply polynomials of $(\tau_i - \tau_j)$, but not of its absolute value.
The effective Hamiltonian used in~\cite{nagai2017self} will be considered in the next section.
At each Monte Carlo step, we insert a new vertex by randomly choosing an imaginary time and spin or remove a vertex from the configuration.
Since we need to calculate the acceptance probability at each step, the weight of the proposed configuration needs to be calculated.
Naively, it will cost $O(n^2 m)$ CPU time for calculating the weight of any configuration $\{(s_i, \tau_i)\}$, but we can do far more than this.

The most naive idea of acceleration is to consider the change of weight by the insertion or removal of a vertex~\cite{nagai2017self}.
For simplicity, we only consider the insertion, while for removal it is almost the same except for the sign.
Suppose we already have a configuration $\{(s_i, \tau_i)\}$ with $n$ vertices ($i = 1, 2, \dots ,n$).
Now we want to insert a new vertex at $\tau$.
For the part of $L$ the change will be
\begin{equation}
    \Delta L = \sum_{i = 1}^{n}(L(\tau_i - \tau) + L(\tau - \tau_i)) + L(0).
\end{equation}
Here, we ignore the overall factor $1 / n$, which can be included back at the end.
As each $L$ term is an order-$m$ polynomial, the total CPU time for calculating $\Delta L$ is $O(n m)$. 
(Note that although we train it as a linear combination of $m$ Chebyshev polynomials, we can combine them into one polynomial after obtaining all coefficients.)
The $J$ part can be done almost the same, except for adding signs, $s_i$ and $s$, for each term.
As for $f(n)$, it can be pre-calculated up to some upper limit since the number of vertices is finite, and if $n$ exceeds the limit, we increase the upper limit to $n$ and calculate the values of the polynomial up to the new upper limit.
Thus, after applying this technique, the total CPU time for one step becomes $O(n m)$.

However, we can do more than this. Consider the structure of $L(\tau_i - \tau)$: it is a polynomial of $(\tau_i - \tau)$, but it can be further expanded and regarded as a polynomial of $\tau$ with $\tau_i$-dependent coefficients.
If $L$ has the following form:
\begin{equation}
    L(x) = \sum_{j = 0}^{m} a_j x^j,
\end{equation}
then 
\begin{align}
  \begin{split}
    L(\tau_i - \tau) &= \sum_{j = 0}^{m} a_j (\tau_i - \tau)^j \\
    &= \sum_{j = 0}^{m} \sum_{k = 0}^j (-1)^k a_j \binom{j}{k}\tau_i^{j - k} \tau^k \\ 
    &= \sum_{k = 0}^{m} \sum_{j = k}^{m} (-1)^k a_j \binom{j}{k}\tau_i^{j - k} \tau^k .
  \end{split}
\end{align}
If we sum over all $n$ vertices of the $L(\tau_i - \tau)$ term, it will be
\begin{equation}
    \sum_{i = 1}^{n}L(\tau_i - \tau) = \sum_{k = 0}^{m} \left(\sum_{i = 1}^{n}\sum_{j = k}^{m} (-1)^k a_j \binom{j}{k}\tau_i^{j - k} \right)\tau^k   .
    \label{eqn:central}
\end{equation}
So, we can store the terms in the parenthesis in Eq.~(\ref{eqn:central}) as a vector of length $(m + 1)$.
The contribution to the vector components from each $\tau_i$ is separable.
This means that we can modify each component with cost of $O(m)$ by performing the summation over $j$ for insertion or deletion of a vertex, and $O(m^2)$ time for the update of all the $(m+1)$ components.
If we have such a vector, the weight difference for inserting a vertex can be calculated by evaluating an order-$m$ polynomial, so the total CPU time cost for one step is $O(m^2)$.

Note that there is a trade-off in this moment pre-calculation method.
Instead of storing the vector, if we rewrite Eq.~(\ref{eqn:central}) as
\begin{align}
  \begin{split}
    \sum_{i = 1}^{n}L(\tau_i - \tau)
    &= \sum_{k = 0}^{m} \left(\sum_{j = 0}^{m - k}\sum_{i = 1}^{n} (-1)^k a_j \binom{j + k}{k}\tau_i^j \right)\tau^k \\ 
    &= \sum_{k = 0}^{m} \left(\sum_{j = 0}^{m - k} (-1)^k a_{j+k} \binom{j + k}{k}\sum_{i = 1}^{n}\tau_i^j \right)\tau^k,
  \end{split}
\end{align}
we can save $(m + 1)$ components of vectors $v_j = \sum_{i = 1}^{n} \tau_i^j$ ($j=1,\cdots,m$).
In this case, each component can be updated in $O(1)$ time for insertion and deletion, so the total cost for updating will be $O(m)$.
However, we need to calculate $m$ components for each vector in the parenthesis for evaluating the weight difference.
The CPU time for this part is $O(m^2)$, and thus the total CPU time will not change: we can choose whether to update faster and calculate the weight difference slower or use a faster calculation of weight difference and spend more time for updating the internal vector.
In both cases, compared with the original $O(nm)$ update, our new method is free from the number of vertices $n \propto \beta U$.
The price is a factor $m$, which is fixed at the beginning and usually small ($m \ll n$), so we can achieve a significant speedup.

One may notice that in Eq.~(\ref{eqn:central}), we have to evaluate the polynomial functions with fixed coefficients many times.
This fact inspires us to do some pre-calculation for the polynomials.
A naive idea is since our $\tau$ can only take values between $0$ and $1$, pre-calculating the function values at $N$ points dividing $[0,1)$ evenly.
Then, for a given $\tau$, we find the interval $[\tau_j, \tau_j + \Delta \tau]$ including $\tau$, and interpolate the function value at $\tau$.
This idea is, in fact, from the interpolation of Green's function for the non-interacting system used in the CT-AUX method.
After the pre-calculation, we can find the function value at any $\tau$ in $O(1)$ time, and the total CPU time is thus reduced to $O(m)$.

The error from this interpolation is $o(\Delta \tau^2)$, where we have another trade-off between memory and error: the more data points we save, the less the error will be.
Since at least we can easily save up to $10^5$ data points,
we can expect the error under control, and this function is also compatible with the algorithm in the next section.
However, we will stop here to consider approximate methods, and in the following, we will consider more rigorous evaluation methods.

\section{Solution for polynomials of absolute value of time difference}

Instead of polynomial of $(\tau_i-\tau_j)$, in the real simulation of fermion systems, a polynomial of $x = 2|\tau_i-\tau_j| - 1$ is used~\cite{Chebyshev0, Chebyshev1, Chebyshev2, Chebyshev3, Chebyshev4} for $L$ (and also for $J$).
This polynomial can be expanded to a polynomial of $|\tau_i-\tau_j|$.
The polynomial of $|\tau_i-\tau_j|$ can be further divided into two parts, each of which is even or odd with respect to $|\tau_i-\tau_j|$.
For the even part, since $|\tau_i-\tau_j|^{2n} = (\tau_i-\tau_j)^{2n}$, no extra effort is needed.
We can apply the technique presented in Sec~\ref{sec:level1} to calculate the weight difference when inserting or removing a vertex.
So, we will focus on the odd part of the polynomial.

First, we consider the simplest case, where $L(\tau_i-\tau) = |\tau_i-\tau|$, so the problem is ``calculating the total distance to all existing vertices from a given vertices at $\tau$.''
We know that
\begin{equation}
    \sum_{i = 1}^{n} |\tau_i - \tau| = \sum_{i = 1}^{n} (\tau_i - \tau) + 2\sum_{i | \tau_i < \tau} (\tau - \tau_i).
\end{equation}
This can be calculated in $O(1)$ time, if we can maintain $\sum_{i = 1}^{n} \tau_i$, $n_l = \sum_{i | \tau_i < \tau}1$ (the number of $\tau_i$ which is smaller than $\tau$), and $\sum_{i | \tau_i < \tau}\tau_i$. The first quantity can be easily maintained, by adding or subtracting $\tau_i$ for each update, so the problem becomes how to deal with the second and the third quantities, related to $\{\tau_i\}$ that are smaller than a given $\tau$.

Formally, we will need a data structure that can do:
\begin{enumerate}
  \item Insert a real number $\tau_i \in [0, 1)$.
  \item Remove a real number $\tau_i$, given its index.
  \item Given a real number $\tau \in [0, 1)$, count the numbers in the data structure that is smaller than $\tau$.
  \item Given a real number $\tau \in [0, 1)$, count the sum of numbers in the data structure that is smaller than $\tau$.
\end{enumerate}
If $\tau$ takes some discrete values (e.g., integers) within some range, instead of real numbers, this is a standard application of the segment tree or the binary indexed tree~\cite{fenwick1994new}.
We can use two binary indexed trees to maintain the number of $\{\tau_i\}$ smaller than $\tau$ and their sum. 

Here, our problem is continuous, so we use another data structure called the \textit{balanced binary search tree}~\cite{knuth1973sorting}.
As its name, the balanced binary search tree is a binary search tree of an ordered set of numbers (both discrete and continuous can be accepted).
Among the family of binary search trees for a single data set, the balanced binary search tree is ``balanced,'' which means the difference of depth for a node's left and right subtrees is very small.
Suppose that we have a degenerating binary search tree, only a chain to the same side (e.g., to the left).
Then, if we want to search for some value to see if it is inside the tree, using the binary search technique, we will need $O(n)$ time to find it or decide that it is not in, where $n$ is the number of data items inside the data structure.
However, if it is balanced, then the maximum depth is $O(\log n)$, so we can find any element in only $O(\log n)$ time.

The balanced binary search tree (from here, we call it \textit{balanced tree} for simplicity) is widely used in many standard libraries and applications.
For example, in the STL of C++, implementation of template data structures (std::set and std::map) is, in fact, by a kind of balanced tree called the \textit{red-black tree}~\cite{RBT1, RBT2}.
This kind of balanced tree is also used at many places in the kernel of Linux OS.
Here in our task, if we can maintain an ordered data structure, finding accumulation properties for values smaller than some given value is expected to be more accessible.
Thus, the balanced tree may be a good choice for our purpose.

Accumulative properties (for example, the sum of all $\tau$'s for a given range) is a natural sort of property to be maintained with binary or general rooted trees since each node can maintain the values of the subtree.
However, what we are interested in here is the ``lower part" of some node, namely, both of the subtree of its left child and the part above it if it is the right child of its parent.
If for our accumulative property both of ``add" and ``subtract" operations are defined, the accumulative property for nodes ``lower" than a given node can be calculated as:
if the given node is the root or left child of its parent, then use the property saved in its left child, otherwise use the total accumulation property of the tree, subtracted by the value saved in its right child (and also subtracted by the value of the node itself if we are asking for a strict ``lower" result).

However, a more natural solution can also work for properties that can only be added but cannot be subtracted.
For example, the minimum value in the lower part of a given node can be ``added'' but not ``subtracted.''
We can maintain the minimum when merging a set of numbers with the minimum of the set, but when removing a subset, only the minimum in that set is not enough since we do not know whether this subset contains all the minimum values or not.
In such a case, we can adopt a sort of random binary search tree called the \textit{splay tree}~\cite{splay}.
Instead of strictly requiring the difference of depth for the two children for each node to be balanced, the splay tree uses a more interesting technique called ``splay" operation, which performs to ``rotate" a node to the root while potentially reducing the imbalance with keeping the binary search tree structure.
Note that the properties of subtrees can be maintained under rotation (or splay) as long as the properties support addition.
The splay operation can be done only in $O(\log n)$ time.
It is proved that this technique can make the complexity for operations on a balanced tree (insertion, deletion, queries for the rank of some value and $k$-th value, and so on) be $O(\log n)$ in average.
As we have seen above, if the node we are considering is the root, then the lower part corresponds to the left subtree, so we can obtain the property lower than some value by first rotating (or ``splaying") the node to the root, then taking the value of its left child.
The average time complexity of the whole operation is $O(\log n)$.
Thus, we see that a balanced tree, or more naturally, a splay tree, can satisfy our four requirements for the data structure and solve the $L(\tau_i-\tau) = |\tau_i-\tau|$ problem with $O(\log n)$ time complexity for both updates and queries.

Now consider a more complex case, i.e., $L(\tau_i-\tau) = |\tau_i-\tau|^3$:
\begin{equation}
    \sum_{i = 1}^{n} |\tau_i - \tau|^3 = \sum_{i = 1}^{n} (\tau_i - \tau)^3 + 2\sum_{i | \tau_i < \tau} (\tau - \tau_i)^3.
\end{equation}
For the first term, we need to store $\sum_i{\tau_i}^k$ for $k = 0,1,2,3$, and then any query and update can be done in $O(1)$ time.
For the second term, we can expand it and maintain $\sum_{i | \tau_i < \tau} \tau_i^k$ ($k = 0,1,2,3$) in the spray tree, so that the difference can still be efficiently calculated. 

From the previous ``trade-off" discussion, we know that if we do not have the balanced tree part, we can either store the coefficients of $\tau^k$ (the complexity for the update is $O(m^2)$ and for the query is $O(m)$) or $\sum_i \tau_i^k$ (the complexity for the update is $O(m)$, but to calculate the weight difference we need $O(m^2)$ time).
Here in the balanced tree part, we will have an extra $\log n$ factor, but we can reduce the total complexity by saving $\sum_i \tau_i^k$: we only need to maintain $O(m)$ properties in the balanced tree, so the update and query for this part are $O(m \log n)$.
Thus, the total CPU time for one step (both update and query) is $O(m^2 + m\log n)$, which is much smaller than $O(mn)$.

\section{Implementation and Benchmark}

\begin{figure}
    \centering
    \includegraphics[width=.9\columnwidth]{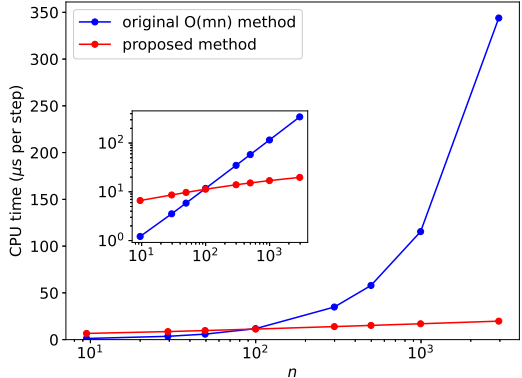}
    \caption{Benchmark of CPU time for the original simple update~\cite{nagai2017self} and our proposed method. Here, $m = 12$ for both $L$ and $J$ polynomials.
We randomly insert or remove vertices while setting the range of the number of vertices as $[n_0 - \sqrt{n_0}, n_0 + \sqrt{n_0}]$ for some pre-chosen $n_0$'s to keep $\langle n\rangle$ at some stable level.
As $n_0$ increases, $\langle n \rangle$, which corresponds to the inverse temperature $\beta$, increases just linearly.
Although the CPU time of the simple update is proportional to $\langle n \rangle$, the CPU time of our proposed method has almost nothing to do with $\langle n \rangle$, only a weak logarithmic dependence for the operations on the splay tree.
Inset shows the same data on a log-log scale.}
    \label{fig:benchmark}
\end{figure}

Finally, we provide a simple benchmark for the SLMC update speed on calculating the weight difference when inserting and removing a vertex (Fig.~\ref{fig:benchmark}).
For the implementation of the method described above, please check our GitHub repository~\cite{github}.
The result is just as we expected: while the CPU time of the simple update proposed in \cite{nagai2017self} is proportional to the average number of vertices, our method proposed in the present paper can calculate the weight difference with the CPU time only weakly depends on $n$. 
The default test for $n = 3000$, $m_{\textrm{c}, J} = m_{\textrm{c}, L} = 12$, $m_{\textrm{c}, f} = 3$ shows a over 10 times speedup compared with the original $O(nm)$ update.

The present benchmark result is consistent with our analysis of the asymptotic time complexity: although for small $n$, the constant factor for the pre-calculation may dominate the CPU time, for large enough $n$, the logarithmic dependence on $n$ of our algorithm achieves an exponential speedup compared with the original algorithm.

\section{Conclusion}

By introducing moment pre-calculation techniques and the help of the splay tree for maintaining the prefix, we have constructed an algorithm that improves the time complexity of a single Monte Carlo step of original SLMC on the CT-AUX method for the single-impurity Anderson model from $O(nm)$ to $O(m^2 + m\log n)$.
Our method gives the same weight as the original one.
For polynomials without absolute values on variables, we also propose an interpolation method that can further improve the time complexity to $O(m)$ with pre-calculation of some polynomial functions. However, we need to consider the error caused by the interpolation process more carefully. 

Although our ideas came from the impurity model and the SLMC algorithm on fermion impurities, the technique itself only depends on more general requirements. As long as the effective Hamiltonian is expressed as a sum of polynomial functions of distances over all pairs of vertices, we can always apply the present technique to obtain an exciting speedup.
It gives a chance for more algorithms to use this idea to improve time complexity, so both save CPU time and improve the accuracy of measurements, which is the main issue, especially in Monte Carlo algorithms facing low-temperature physics.
We are also looking forward to more discussions about applying the current technique and welcome further improvements based on this work.

\section*{Acknowlegements}

The authors thank T. Kato, H. Suwa, and Y. Nagai for fruitful discussions and comments on algorithms and applications. R.C. was supported by the Global Science Graduate Course (GSGC) program of The University of Tokyo. This work was partially supported by JSPS KAKENHI (No.~17K05564 and 20H01824).

\bibliography{fastSLMC}

\begin{thebibliography}{27}%
\makeatletter
\providecommand \@ifxundefined [1]{%
 \@ifx{#1\undefined}
}%
\providecommand \@ifnum [1]{%
 \ifnum #1\expandafter \@firstoftwo
 \else \expandafter \@secondoftwo
 \fi
}%
\providecommand \@ifx [1]{%
 \ifx #1\expandafter \@firstoftwo
 \else \expandafter \@secondoftwo
 \fi
}%
\providecommand \natexlab [1]{#1}%
\providecommand \enquote  [1]{``#1''}%
\providecommand \bibnamefont  [1]{#1}%
\providecommand \bibfnamefont [1]{#1}%
\providecommand \citenamefont [1]{#1}%
\providecommand \href@noop [0]{\@secondoftwo}%
\providecommand \href [0]{\begingroup \@sanitize@url \@href}%
\providecommand \@href[1]{\@@startlink{#1}\@@href}%
\providecommand \@@href[1]{\endgroup#1\@@endlink}%
\providecommand \@sanitize@url [0]{\catcode `\\12\catcode `\$12\catcode
  `\&12\catcode `\#12\catcode `\^12\catcode `\_12\catcode `\%12\relax}%
\providecommand \@@startlink[1]{}%
\providecommand \@@endlink[0]{}%
\providecommand \url  [0]{\begingroup\@sanitize@url \@url }%
\providecommand \@url [1]{\endgroup\@href {#1}{\urlprefix }}%
\providecommand \urlprefix  [0]{URL }%
\providecommand \Eprint [0]{\href }%
\providecommand \doibase [0]{https://doi.org/}%
\providecommand \selectlanguage [0]{\@gobble}%
\providecommand \bibinfo  [0]{\@secondoftwo}%
\providecommand \bibfield  [0]{\@secondoftwo}%
\providecommand \translation [1]{[#1]}%
\providecommand \BibitemOpen [0]{}%
\providecommand \bibitemStop [0]{}%
\providecommand \bibitemNoStop [0]{.\EOS\space}%
\providecommand \EOS [0]{\spacefactor3000\relax}%
\providecommand \BibitemShut  [1]{\csname bibitem#1\endcsname}%
\let\auto@bib@innerbib\@empty
%</preamble>
\bibitem [{\citenamefont {Landau}\ and\ \citenamefont
  {Binder}(2014)}]{LandauB2014}%
  \BibitemOpen
  \bibfield  {author} {\bibinfo {author} {\bibfnamefont {D.~P.}\ \bibnamefont
  {Landau}}\ and\ \bibinfo {author} {\bibfnamefont {K.}~\bibnamefont
  {Binder}},\ }\href@noop {} {\emph {\bibinfo {title} {A Guide to {Monte}
  {Carlo} Simulations in Statistical Physics}}},\ \bibinfo {edition} {4th}\
  ed.\ (\bibinfo  {publisher} {Cambridge University Press},\ \bibinfo {address}
  {Cambridge},\ \bibinfo {year} {2014})\BibitemShut {NoStop}%
\bibitem [{\citenamefont {Gubernatis}\ \emph {et~al.}(2016)\citenamefont
  {Gubernatis}, \citenamefont {Kawashima},\ and\ \citenamefont
  {Werner}}]{GubernatisKW2016}%
  \BibitemOpen
  \bibfield  {author} {\bibinfo {author} {\bibfnamefont {J.}~\bibnamefont
  {Gubernatis}}, \bibinfo {author} {\bibfnamefont {N.}~\bibnamefont
  {Kawashima}},\ and\ \bibinfo {author} {\bibfnamefont {P.}~\bibnamefont
  {Werner}},\ }\href@noop {} {\emph {\bibinfo {title} {Quantum {Monte} {Carlo}
  Methods: Algorithms for Lattice Models}}}\ (\bibinfo  {publisher} {Cambridge
  University Press},\ \bibinfo {address} {Cambridge},\ \bibinfo {year}
  {2016})\BibitemShut {NoStop}%
\bibitem [{\citenamefont {Rombouts}\ \emph {et~al.}(1998)\citenamefont
  {Rombouts}, \citenamefont {Heyde},\ and\ \citenamefont
  {Jachowicz}}]{CT-AUX-1}%
  \BibitemOpen
  \bibfield  {author} {\bibinfo {author} {\bibfnamefont {S.}~\bibnamefont
  {Rombouts}}, \bibinfo {author} {\bibfnamefont {K.}~\bibnamefont {Heyde}},\
  and\ \bibinfo {author} {\bibfnamefont {N.}~\bibnamefont {Jachowicz}},\
  }\bibfield  {title} {\bibinfo {title} {A discrete {Hubbard}-{Stratonovich}
  decomposition for general, fermionic two-body interactions},\ }\href@noop {}
  {\bibfield  {journal} {\bibinfo  {journal} {Phys. Lett. A}\ }\textbf
  {\bibinfo {volume} {242}},\ \bibinfo {pages} {271} (\bibinfo {year}
  {1998})}\BibitemShut {NoStop}%
\bibitem [{\citenamefont {Rombouts}\ \emph {et~al.}(1999)\citenamefont
  {Rombouts}, \citenamefont {Heyde},\ and\ \citenamefont
  {Jachowicz}}]{CT-AUX0}%
  \BibitemOpen
  \bibfield  {author} {\bibinfo {author} {\bibfnamefont {S.}~\bibnamefont
  {Rombouts}}, \bibinfo {author} {\bibfnamefont {K.}~\bibnamefont {Heyde}},\
  and\ \bibinfo {author} {\bibfnamefont {N.}~\bibnamefont {Jachowicz}},\
  }\bibfield  {title} {\bibinfo {title} {Quantum {Monte} {Carlo} method for
  fermions, free of discretization errors},\ }\href@noop {} {\bibfield
  {journal} {\bibinfo  {journal} {Phys. Rev. Lett.}\ }\textbf {\bibinfo
  {volume} {82}},\ \bibinfo {pages} {4155} (\bibinfo {year}
  {1999})}\BibitemShut {NoStop}%
\bibitem [{\citenamefont {Gull}\ \emph {et~al.}(2008)\citenamefont {Gull},
  \citenamefont {Werner}, \citenamefont {Parcollet},\ and\ \citenamefont
  {Troyer}}]{CT-AUX1}%
  \BibitemOpen
  \bibfield  {author} {\bibinfo {author} {\bibfnamefont {E.}~\bibnamefont
  {Gull}}, \bibinfo {author} {\bibfnamefont {P.}~\bibnamefont {Werner}},
  \bibinfo {author} {\bibfnamefont {O.}~\bibnamefont {Parcollet}},\ and\
  \bibinfo {author} {\bibfnamefont {M.}~\bibnamefont {Troyer}},\ }\bibfield
  {title} {\bibinfo {title} {Continuous-time auxiliary-field {Monte} {Carlo}
  for quantum impurity models},\ }\href@noop {} {\bibfield  {journal} {\bibinfo
   {journal} {Euro. Phys. Lett.}\ }\textbf {\bibinfo {volume} {82}},\ \bibinfo
  {pages} {57003} (\bibinfo {year} {2008})}\BibitemShut {NoStop}%
\bibitem [{\citenamefont {Gull}\ \emph
  {et~al.}(2011{\natexlab{a}})\citenamefont {Gull}, \citenamefont {Millis},
  \citenamefont {Lichtenstein}, \citenamefont {Rubtsov}, \citenamefont
  {Troyer},\ and\ \citenamefont {Werner}}]{CTAlgorithmsGull}%
  \BibitemOpen
  \bibfield  {author} {\bibinfo {author} {\bibfnamefont {E.}~\bibnamefont
  {Gull}}, \bibinfo {author} {\bibfnamefont {A.~J.}\ \bibnamefont {Millis}},
  \bibinfo {author} {\bibfnamefont {A.~I.}\ \bibnamefont {Lichtenstein}},
  \bibinfo {author} {\bibfnamefont {A.~N.}\ \bibnamefont {Rubtsov}}, \bibinfo
  {author} {\bibfnamefont {M.}~\bibnamefont {Troyer}},\ and\ \bibinfo {author}
  {\bibfnamefont {P.}~\bibnamefont {Werner}},\ }\bibfield  {title} {\bibinfo
  {title} {Continuous-time {Monte} {Carlo} methods for quantum impurity
  models},\ }\href@noop {} {\bibfield  {journal} {\bibinfo  {journal} {Rev.
  Mod. Phys.}\ }\textbf {\bibinfo {volume} {83}},\ \bibinfo {pages} {349}
  (\bibinfo {year} {2011}{\natexlab{a}})}\BibitemShut {NoStop}%
\bibitem [{\citenamefont {Anderson}(1961)}]{anderson1961localized}%
  \BibitemOpen
  \bibfield  {author} {\bibinfo {author} {\bibfnamefont {P.~W.}\ \bibnamefont
  {Anderson}},\ }\bibfield  {title} {\bibinfo {title} {Localized magnetic
  states in metals},\ }\href@noop {} {\bibfield  {journal} {\bibinfo  {journal}
  {Phys. Rev.}\ }\textbf {\bibinfo {volume} {124}},\ \bibinfo {pages} {41}
  (\bibinfo {year} {1961})}\BibitemShut {NoStop}%
\bibitem [{\citenamefont {Nagai}\ \emph {et~al.}(2017)\citenamefont {Nagai},
  \citenamefont {Shen}, \citenamefont {Qi}, \citenamefont {Liu},\ and\
  \citenamefont {Fu}}]{nagai2017self}%
  \BibitemOpen
  \bibfield  {author} {\bibinfo {author} {\bibfnamefont {Y.}~\bibnamefont
  {Nagai}}, \bibinfo {author} {\bibfnamefont {H.}~\bibnamefont {Shen}},
  \bibinfo {author} {\bibfnamefont {Y.}~\bibnamefont {Qi}}, \bibinfo {author}
  {\bibfnamefont {J.}~\bibnamefont {Liu}},\ and\ \bibinfo {author}
  {\bibfnamefont {L.}~\bibnamefont {Fu}},\ }\bibfield  {title} {\bibinfo
  {title} {Self-learning {Monte} {Carlo} method: Continuous-time algorithm},\
  }\href@noop {} {\bibfield  {journal} {\bibinfo  {journal} {Phys. Rev. B}\
  }\textbf {\bibinfo {volume} {96}},\ \bibinfo {pages} {161102} (\bibinfo
  {year} {2017})}\BibitemShut {NoStop}%
\bibitem [{\citenamefont {Blankenbecler}\ \emph {et~al.}(1981)\citenamefont
  {Blankenbecler}, \citenamefont {Scalapino},\ and\ \citenamefont
  {Sugar}}]{model0}%
  \BibitemOpen
  \bibfield  {author} {\bibinfo {author} {\bibfnamefont {R.}~\bibnamefont
  {Blankenbecler}}, \bibinfo {author} {\bibfnamefont {D.~J.}\ \bibnamefont
  {Scalapino}},\ and\ \bibinfo {author} {\bibfnamefont {R.~L.}\ \bibnamefont
  {Sugar}},\ }\bibfield  {title} {\bibinfo {title} {{Monte} {Carlo}
  calculations of coupled boson-fermion systems. {I}},\ }\href@noop {}
  {\bibfield  {journal} {\bibinfo  {journal} {Phys. Rev. D}\ }\textbf {\bibinfo
  {volume} {24}},\ \bibinfo {pages} {2278} (\bibinfo {year}
  {1981})}\BibitemShut {NoStop}%
\bibitem [{\citenamefont {Hirsch}\ and\ \citenamefont {Fye}(1986)}]{model1}%
  \BibitemOpen
  \bibfield  {author} {\bibinfo {author} {\bibfnamefont {J.~E.}\ \bibnamefont
  {Hirsch}}\ and\ \bibinfo {author} {\bibfnamefont {R.~M.}\ \bibnamefont
  {Fye}},\ }\bibfield  {title} {\bibinfo {title} {{Monte} {Carlo} method for
  magnetic impurities in metals},\ }\href@noop {} {\bibfield  {journal}
  {\bibinfo  {journal} {Phys. Rev. Lett.}\ }\textbf {\bibinfo {volume} {56}},\
  \bibinfo {pages} {2521} (\bibinfo {year} {1986})}\BibitemShut {NoStop}%
\bibitem [{\citenamefont {Gull}\ \emph
  {et~al.}(2011{\natexlab{b}})\citenamefont {Gull}, \citenamefont {Staar},
  \citenamefont {Fuchs}, \citenamefont {Nukala}, \citenamefont {Summers},
  \citenamefont {Pruschke}, \citenamefont {Schulthess},\ and\ \citenamefont
  {Maier}}]{gullfastupdate}%
  \BibitemOpen
  \bibfield  {author} {\bibinfo {author} {\bibfnamefont {E.}~\bibnamefont
  {Gull}}, \bibinfo {author} {\bibfnamefont {P.}~\bibnamefont {Staar}},
  \bibinfo {author} {\bibfnamefont {S.}~\bibnamefont {Fuchs}}, \bibinfo
  {author} {\bibfnamefont {P.}~\bibnamefont {Nukala}}, \bibinfo {author}
  {\bibfnamefont {M.~S.}\ \bibnamefont {Summers}}, \bibinfo {author}
  {\bibfnamefont {T.}~\bibnamefont {Pruschke}}, \bibinfo {author}
  {\bibfnamefont {T.~C.}\ \bibnamefont {Schulthess}},\ and\ \bibinfo {author}
  {\bibfnamefont {T.}~\bibnamefont {Maier}},\ }\bibfield  {title} {\bibinfo
  {title} {Submatrix updates for the continuous-time auxiliary-field
  algorithm},\ }\href@noop {} {\bibfield  {journal} {\bibinfo  {journal} {Phys.
  Rev. B}\ }\textbf {\bibinfo {volume} {83}},\ \bibinfo {pages} {075122}
  (\bibinfo {year} {2011}{\natexlab{b}})}\BibitemShut {NoStop}%
\bibitem [{\citenamefont {Liu}\ \emph {et~al.}(2017{\natexlab{a}})\citenamefont
  {Liu}, \citenamefont {Qi}, \citenamefont {Meng},\ and\ \citenamefont
  {Fu}}]{SLMC0}%
  \BibitemOpen
  \bibfield  {author} {\bibinfo {author} {\bibfnamefont {J.}~\bibnamefont
  {Liu}}, \bibinfo {author} {\bibfnamefont {Y.}~\bibnamefont {Qi}}, \bibinfo
  {author} {\bibfnamefont {Z.~Y.}\ \bibnamefont {Meng}},\ and\ \bibinfo
  {author} {\bibfnamefont {L.}~\bibnamefont {Fu}},\ }\bibfield  {title}
  {\bibinfo {title} {Self-learning {Monte} {Carlo} method},\ }\href@noop {}
  {\bibfield  {journal} {\bibinfo  {journal} {Phys. Rev. B}\ }\textbf {\bibinfo
  {volume} {95}},\ \bibinfo {pages} {041101} (\bibinfo {year}
  {2017}{\natexlab{a}})}\BibitemShut {NoStop}%
\bibitem [{\citenamefont {Liu}\ \emph {et~al.}(2017{\natexlab{b}})\citenamefont
  {Liu}, \citenamefont {Shen}, \citenamefont {Qi}, \citenamefont {Meng},\ and\
  \citenamefont {Fu}}]{SLMC1}%
  \BibitemOpen
  \bibfield  {author} {\bibinfo {author} {\bibfnamefont {J.}~\bibnamefont
  {Liu}}, \bibinfo {author} {\bibfnamefont {H.}~\bibnamefont {Shen}}, \bibinfo
  {author} {\bibfnamefont {Y.}~\bibnamefont {Qi}}, \bibinfo {author}
  {\bibfnamefont {Z.~Y.}\ \bibnamefont {Meng}},\ and\ \bibinfo {author}
  {\bibfnamefont {L.}~\bibnamefont {Fu}},\ }\bibfield  {title} {\bibinfo
  {title} {Self-learning {Monte} {Carlo} method and cumulative update in
  fermion systems},\ }\href@noop {} {\bibfield  {journal} {\bibinfo  {journal}
  {Phys. Rev. B}\ }\textbf {\bibinfo {volume} {95}},\ \bibinfo {pages} {241104}
  (\bibinfo {year} {2017}{\natexlab{b}})}\BibitemShut {NoStop}%
\bibitem [{\citenamefont {Xu}\ \emph {et~al.}(2017)\citenamefont {Xu},
  \citenamefont {Qi}, \citenamefont {Liu}, \citenamefont {Fu},\ and\
  \citenamefont {Meng}}]{SLMC2}%
  \BibitemOpen
  \bibfield  {author} {\bibinfo {author} {\bibfnamefont {X.~Y.}\ \bibnamefont
  {Xu}}, \bibinfo {author} {\bibfnamefont {Y.}~\bibnamefont {Qi}}, \bibinfo
  {author} {\bibfnamefont {J.}~\bibnamefont {Liu}}, \bibinfo {author}
  {\bibfnamefont {L.}~\bibnamefont {Fu}},\ and\ \bibinfo {author}
  {\bibfnamefont {Z.~Y.}\ \bibnamefont {Meng}},\ }\bibfield  {title} {\bibinfo
  {title} {Self-learning quantum {Monte} {Carlo} method in interacting fermion
  systems},\ }\href@noop {} {\bibfield  {journal} {\bibinfo  {journal} {Phys.
  Rev. B}\ }\textbf {\bibinfo {volume} {96}},\ \bibinfo {pages} {041119(R)}
  (\bibinfo {year} {2017})}\BibitemShut {NoStop}%
\bibitem [{\citenamefont {Nagai}\ \emph {et~al.}(2020)\citenamefont {Nagai},
  \citenamefont {Okumura},\ and\ \citenamefont {Tanaka}}]{nagai2020self}%
  \BibitemOpen
  \bibfield  {author} {\bibinfo {author} {\bibfnamefont {Y.}~\bibnamefont
  {Nagai}}, \bibinfo {author} {\bibfnamefont {M.}~\bibnamefont {Okumura}},\
  and\ \bibinfo {author} {\bibfnamefont {A.}~\bibnamefont {Tanaka}},\
  }\bibfield  {title} {\bibinfo {title} {Self-learning {Monte} {Carlo} method
  with {Behler}-{Parrinello} neural networks},\ }\href@noop {} {\bibfield
  {journal} {\bibinfo  {journal} {Phys. Rev. B}\ }\textbf {\bibinfo {volume}
  {101}},\ \bibinfo {pages} {115111} (\bibinfo {year} {2020})}\BibitemShut
  {NoStop}%
\bibitem [{\citenamefont {Yamamoto}\ \emph {et~al.}(2018)\citenamefont
  {Yamamoto}, \citenamefont {Kato}, \citenamefont {Kato},\ and\ \citenamefont
  {Saito}}]{yamamoto2018heat}%
  \BibitemOpen
  \bibfield  {author} {\bibinfo {author} {\bibfnamefont {T.}~\bibnamefont
  {Yamamoto}}, \bibinfo {author} {\bibfnamefont {M.}~\bibnamefont {Kato}},
  \bibinfo {author} {\bibfnamefont {T.}~\bibnamefont {Kato}},\ and\ \bibinfo
  {author} {\bibfnamefont {K.}~\bibnamefont {Saito}},\ }\bibfield  {title}
  {\bibinfo {title} {Heat transport via a local two-state system near thermal
  equilibrium},\ }\href@noop {} {\bibfield  {journal} {\bibinfo  {journal} {New
  J. Phys.}\ }\textbf {\bibinfo {volume} {20}},\ \bibinfo {pages} {093014}
  (\bibinfo {year} {2018})}\BibitemShut {NoStop}%
\bibitem [{\citenamefont {Wolf}\ \emph {et~al.}(2014)\citenamefont {Wolf},
  \citenamefont {McCulloch}, \citenamefont {Parcollet},\ and\ \citenamefont
  {Schollw{\"o}ck}}]{Chebyshev0}%
  \BibitemOpen
  \bibfield  {author} {\bibinfo {author} {\bibfnamefont {F.~A.}\ \bibnamefont
  {Wolf}}, \bibinfo {author} {\bibfnamefont {I.~P.}\ \bibnamefont {McCulloch}},
  \bibinfo {author} {\bibfnamefont {O.}~\bibnamefont {Parcollet}},\ and\
  \bibinfo {author} {\bibfnamefont {U.}~\bibnamefont {Schollw{\"o}ck}},\
  }\bibfield  {title} {\bibinfo {title} {Chebyshev matrix product state
  impurity solver for dynamical mean-field theory},\ }\href@noop {} {\bibfield
  {journal} {\bibinfo  {journal} {Phys. Rev. B}\ }\textbf {\bibinfo {volume}
  {90}},\ \bibinfo {pages} {115124} (\bibinfo {year} {2014})}\BibitemShut
  {NoStop}%
\bibitem [{\citenamefont {Braun}\ and\ \citenamefont
  {Schmitteckert}(2014)}]{Chebyshev1}%
  \BibitemOpen
  \bibfield  {author} {\bibinfo {author} {\bibfnamefont {A.}~\bibnamefont
  {Braun}}\ and\ \bibinfo {author} {\bibfnamefont {P.}~\bibnamefont
  {Schmitteckert}},\ }\bibfield  {title} {\bibinfo {title} {Numerical
  evaluation of {Green's} functions based on the {Chebyshev} expansion},\
  }\href@noop {} {\bibfield  {journal} {\bibinfo  {journal} {Phys. Rev. B}\
  }\textbf {\bibinfo {volume} {90}},\ \bibinfo {pages} {165112} (\bibinfo
  {year} {2014})}\BibitemShut {NoStop}%
\bibitem [{\citenamefont {Covaci}\ \emph {et~al.}(2010)\citenamefont {Covaci},
  \citenamefont {Peeters},\ and\ \citenamefont {Berciu}}]{Chebyshev2}%
  \BibitemOpen
  \bibfield  {author} {\bibinfo {author} {\bibfnamefont {L.}~\bibnamefont
  {Covaci}}, \bibinfo {author} {\bibfnamefont {F.}~\bibnamefont {Peeters}},\
  and\ \bibinfo {author} {\bibfnamefont {M.}~\bibnamefont {Berciu}},\
  }\bibfield  {title} {\bibinfo {title} {Efficient numerical approach to
  inhomogeneous superconductivity: the {Chebyshev}-{Bogoliubov}-de {Gennes}
  method},\ }\href@noop {} {\bibfield  {journal} {\bibinfo  {journal} {Phys.
  Rev. Lett.}\ }\textbf {\bibinfo {volume} {105}},\ \bibinfo {pages} {167006}
  (\bibinfo {year} {2010})}\BibitemShut {NoStop}%
\bibitem [{\citenamefont {Nagai}\ \emph {et~al.}(2012)\citenamefont {Nagai},
  \citenamefont {Ota},\ and\ \citenamefont {Machida}}]{Chebyshev3}%
  \BibitemOpen
  \bibfield  {author} {\bibinfo {author} {\bibfnamefont {Y.}~\bibnamefont
  {Nagai}}, \bibinfo {author} {\bibfnamefont {Y.}~\bibnamefont {Ota}},\ and\
  \bibinfo {author} {\bibfnamefont {M.}~\bibnamefont {Machida}},\ }\bibfield
  {title} {\bibinfo {title} {Efficient numerical self-consistent mean-field
  approach for fermionic many-body systems by polynomial expansion on spectral
  density},\ }\href@noop {} {\bibfield  {journal} {\bibinfo  {journal} {J.
  Phys. Soc. Japan}\ }\textbf {\bibinfo {volume} {81}},\ \bibinfo {pages}
  {024710} (\bibinfo {year} {2012})}\BibitemShut {NoStop}%
\bibitem [{\citenamefont {Sota}\ and\ \citenamefont
  {Tohyama}(2008)}]{Chebyshev4}%
  \BibitemOpen
  \bibfield  {author} {\bibinfo {author} {\bibfnamefont {S.}~\bibnamefont
  {Sota}}\ and\ \bibinfo {author} {\bibfnamefont {T.}~\bibnamefont {Tohyama}},\
  }\bibfield  {title} {\bibinfo {title} {Low-temperature density matrix
  renormalization group using regulated polynomial expansion},\ }\href@noop {}
  {\bibfield  {journal} {\bibinfo  {journal} {Phys. Rev. B}\ }\textbf {\bibinfo
  {volume} {78}},\ \bibinfo {pages} {113101} (\bibinfo {year}
  {2008})}\BibitemShut {NoStop}%
\bibitem [{\citenamefont {Fenwick}(1994)}]{fenwick1994new}%
  \BibitemOpen
  \bibfield  {author} {\bibinfo {author} {\bibfnamefont {P.~M.}\ \bibnamefont
  {Fenwick}},\ }\bibfield  {title} {\bibinfo {title} {A new data structure for
  cumulative frequency tables},\ }\href@noop {} {\bibfield  {journal} {\bibinfo
   {journal} {Software Pract. Exper.}\ }\textbf {\bibinfo {volume} {24}},\
  \bibinfo {pages} {327} (\bibinfo {year} {1994})}\BibitemShut {NoStop}%
\bibitem [{\citenamefont {Knuth}(1973)}]{knuth1973sorting}%
  \BibitemOpen
  \bibfield  {author} {\bibinfo {author} {\bibfnamefont {D.}~\bibnamefont
  {Knuth}},\ }\href@noop {} {\emph {\bibinfo {title} {The Art Of Computer
  Programming, vol. 3: Sorting And Searching}}}\ (\bibinfo  {publisher}
  {Addison-Wesley},\ \bibinfo {year} {1973})\BibitemShut {NoStop}%
\bibitem [{\citenamefont {Bayer}(1972)}]{RBT1}%
  \BibitemOpen
  \bibfield  {author} {\bibinfo {author} {\bibfnamefont {R.}~\bibnamefont
  {Bayer}},\ }\bibfield  {title} {\bibinfo {title} {Symmetric binary b-trees:
  Data structure and maintenance algorithms},\ }\href@noop {} {\bibfield
  {journal} {\bibinfo  {journal} {Acta Informatica}\ }\textbf {\bibinfo
  {volume} {1}},\ \bibinfo {pages} {290} (\bibinfo {year} {1972})}\BibitemShut
  {NoStop}%
\bibitem [{\citenamefont {Guibas}\ and\ \citenamefont
  {Sedgewick}(1978)}]{RBT2}%
  \BibitemOpen
  \bibfield  {author} {\bibinfo {author} {\bibfnamefont {L.~J.}\ \bibnamefont
  {Guibas}}\ and\ \bibinfo {author} {\bibfnamefont {R.}~\bibnamefont
  {Sedgewick}},\ }\bibfield  {title} {\bibinfo {title} {A dichromatic framework
  for balanced trees},\ }in\ \href@noop {} {\emph {\bibinfo {booktitle} {19th
  Annual Symposium on Foundations of Computer Science (sfcs 1978)}}}\ (\bibinfo
  {organization} {IEEE},\ \bibinfo {year} {1978})\ pp.\ \bibinfo {pages}
  {8--21}\BibitemShut {NoStop}%
\bibitem [{\citenamefont {Sleator}\ and\ \citenamefont {Tarjan}(1985)}]{splay}%
  \BibitemOpen
  \bibfield  {author} {\bibinfo {author} {\bibfnamefont {D.~D.}\ \bibnamefont
  {Sleator}}\ and\ \bibinfo {author} {\bibfnamefont {R.~E.}\ \bibnamefont
  {Tarjan}},\ }\bibfield  {title} {\bibinfo {title} {Self-adjusting binary
  search trees},\ }\href@noop {} {\bibfield  {journal} {\bibinfo  {journal} {J.
  ACM}\ }\textbf {\bibinfo {volume} {32}},\ \bibinfo {pages} {652} (\bibinfo
  {year} {1985})}\BibitemShut {NoStop}%
\bibitem [{git()}]{github}%
  \BibitemOpen
  \href@noop {} {}\bibinfo {howpublished}
  {\url{https://github.com/CaoRX/SLMC-demo}}\BibitemShut {NoStop}%
\end{thebibliography}%

\end{document}